\documentstyle[12pt]{article}

\newcommand{\bce}{\begin{center}}
\newcommand{\ece}{\end{center}}
\newcommand{\beq}{\begin{equation}}
\newcommand{\eeq}{\end{equation}}
\newcommand{\bea}{\vspace{0.25cm}\begin{eqnarray}}
\newcommand{\eea}{\end{eqnarray}}

\newcommand{\ba}{\begin{array}}
\newcommand{\ea}{\end{array}}

\newcommand{\doublespace}{
    \renewcommand{\baselinestretch}{1.6}\large\normalsize}

\def\lsim{\mathrel{\rlap{\lower4pt\hbox{\hskip1pt$\sim$}}
    \raise1pt\hbox{$<$}}}	  
\def\gsim{\mathrel{\rlap{\lower4pt\hbox{\hskip1pt$\sim$}}
    \raise1pt\hbox{$>$}}}	  

\def\Pom{{\bf I\!P}}

  \advance\oddsidemargin  by -1in
  \advance\evensidemargin by -1in
  \advance\topmargin      by -0.5in
\def\lsim{\mathrel{\rlap{\lower4pt\hbox{\hskip1pt$\sim$}}
    \raise1pt\hbox{$<$}}}         
\def\gsim{\mathrel{\rlap{\lower4pt\hbox{\hskip1pt$\sim$}}
    \raise1pt\hbox{$>$}}}         

\def\Pom{{\bf I\!P}}
\def\beq{\begin{equation}}
\def\endeq{\end{equation}}
\def\arr{\begin{eqnarray}}
\def\endarr{\end{eqnarray}}
\makeindex

 \doublespace

\begin{document}
\phantom{.}{\bf \Large \hspace{10.0cm} KFA-IKP(Th)-1994-10\\ }
\phantom{.}\hspace{10.9cm}28 February 1994\\

\begin{center}
{\bf \huge How to measure the intercept\\
of the BFKL pomeron at HERA
\vspace{0.5cm}\\}
{\bf \Large
N.N.~Nikolaev$^{a,b}$,
and B.G.~Zakharov$^{b,c}$ \vspace{0.3cm} \\}
{\it
$^{a}$IKP(Theorie), KFA J{\"u}lich, 5170 J{\"u}lich, Germany
\medskip\\
$^{b}$L. D. Landau Institute for Theoretical Physics, GSP-1,
117940, \\
ul. Kosygina 2, Moscow V-334, Russia.
\medskip\\
$^{c}$Interdisciplinary Laboratory for Advance Studies (ILAS),\\
Miramare, I-34014 Trieste, Italy
\vspace{1.0cm}\\
}
{\LARGE
Abstract}\\
\end{center}
Determination of the intercept of the BFKL pomeron is one
of the pressing issues in the high energy physics.
Earlier we have shown that,
at the dipole size $r=r_{\Delta}=(0.1-0.2)$f,
the dipole cross section
$\sigma(x,r)$  which is a solution of the generalized BFKL
equation, exhibits
a precocious asymptotic behavior
$\sigma(x,r)\propto \left({1\over x}\right)^{\Delta_{\Pom}}$.
In this paper we discuss how measuring $F_{L}(x,Q^{2})$ and
$\partial F_{T}(x,Q^{2})/\partial \log Q^{2}$ at
$Q^{2}= (10-40)$GeV$^{2}$ and $Q^{2}=(2-10)$GeV$^{2}$,
respectively, one can probe $\sigma(x,r_{\Delta})$ and
directly determine the intercept $\Delta_{\Pom}$ of the
BFKL pomeron in the HERA experiments.

\begin{center}
Submitted to {\sl Physics Letters B}\\
E-mail: kph154@zam001.zam.kfa-juelich.de
\end{center}

\pagebreak

The intercept $\Delta_{\Pom}$ of the BFKL pomeron [1] is one of
the fundamental parametes of QCD. In the BFKL regime, at
sufficiently large ${1\over x}$, all the
structure functions must have the behaviour
\beq
F_{i}(x,Q^{2}) \propto \left({1\over x}\right)^{-\Delta_{\Pom}}
\label{eq:1}
\endeq
with the
exponent (intercept) $\Delta_{\Pom}$ which is independent of
$Q^{2}$ (hereafter $Q^{2}$ is the virtuality of the photon,
$x$ is the Bjorken variable and we refer to deep inelastic
scattering (DIS) at very large ${1\over x}$ as the diffractive
DIS). Very large ${1\over x}\sim 10^{5}$ are attainable in the HERA
DIS experiments, and it is important to find out whether the
pomeron's intercept $\Delta_{\Pom}$ can be measured at HERA or not.

Recently we have examined [2-4] the onset of the BFKL regime
in the framework of our $s$-channel lightcone approach to
diffractive DIS [5,6] and our generalized BFKL equation for
the dipole cross section [2-4,6,7]. Our conclusion was that the
the onset of the BFKL asymptotics as a global feature of
DIS is unlikely to happen in the kinematic range of
HERA. However, an important observation [2,3] is that the
$x$-dependence of the dipole cross section $\sigma(x,r)$,
and which is a solution of our generalized BFKL equation,
exhibits for the dipole size $r=r_{\Delta}=(0.1-0.2)$f,
a precocious asymptotic behaviour (\ref{eq:1}).
This dictates the obvious
strategy: in order to expertimentally determine the pomeron's
intercept $\Delta_{\Pom}$,
one must find observables which are dominated by
$\sigma(x,r)$ at $r\sim r_{\Delta}$. An example
of such an observable  - the real and virtual photoproduction
of the open charm at $Q^{2}\lsim 10$GeV$^{2}$ - has already been
discussed in [2,3].

In this paper we demonstrate that the longitudinal structure
function $F_{L}(x,Q^{2})$ and the slope of the transverse
structure function $\partial F_{T}(x,Q^{2})/\partial \log Q^{2}$
emerge as local probes of the dipole cross section at
$
r^{2}\approx {B_{T,L}/ Q^{2}}
$
with $B_{T}\approx 2.3  $ and $B_{L}\approx 11$. Therefore,
in order to measure the pomeron's
intercept $\Delta_{\Pom}$, one must concentrate on the
$x$-dependence of $F_{L}(x,Q^{2})$ at
\beq
Q^{2}={B_{L}\over r_{\Delta}^{2}} \approx (10-40){\rm GeV}^{2}
\label{eq:3}
\endeq
and of $\partial F_{T}(x,Q^{2})/\partial
\log Q^{2}$ at
\beq
Q^{2}={B_{T}\over r_{\Delta}^{2}} \approx (2-10){\rm GeV}^{2} \, .
\label{eq:4}
\endeq
This range of $Q^{2}$ is easily accessible at HERA and, as a
matter of fact, falls in the region of highest counting rates,
which makes the measurement of $\Delta_{\Pom}$ possible already
in the near future.

The starting point of derivation of the above results is the
$s$-channel lightcone approach to diffractive DIS [4,5]. Here
the photoabsorption cross section for the
(T) transverse and (L) longitudinal photons can be written down
as
\beq
\sigma_{T,L}(\gamma^{*}N,x,Q^{2})=
\int_{0}^{1} dz\int d^{2}\vec{r}\,\,
\vert\Psi_{T,L}(z,r)\vert^{2}\sigma(x,r)\,\,,
\label{eq:5}
\endeq
where $\sigma(x,r)$ is the total cross section of interaction
of the $q\bar{q}$ colour dipole of transverse size $r$ with
the nucleon target, and is a solution of the
generalized BFKL equation [2-4,6,7].
The wave functions of the $q\bar{q}$ Fock states of the photon were
derived in [5] and read
\beq
\vert\Psi_{T}(z,r)\vert^{2}=
\sum e_{f}^{2}|\Psi_{T}^{(f\bar{f})}(z,r)|^{2}=
{6\alpha_{em} \over (2\pi)^{2}}
\sum_{1}^{N_{f}}Z_{f}^{2}
\{[z^{2}+(1-z)^{2}]\varepsilon^{2}K_{1}(\varepsilon r)^{2}+
m_{f}^{2}K_{0}(\varepsilon r)^{2}\}\,\,,
\label{eq:6}
\endeq
\beq
\vert\Psi_{L}(z,r)\vert^{2}=
\sum e_{f}^{2}|\Psi_{T}^{(f\bar{f})}(z,r)|^{2}=
{6\alpha_{em} \over (2\pi)^{2}}
\sum_{1}^{N_{f}}4Z_{f}^{2}\,\,
Q^{2}\,z^{2}(1-z)^{2}K_{0}(\varepsilon r)^{2}\,\,,
\label{eq:7}
\endeq
where $K_{\nu}(x)$ are the modified Bessel functions,
$\varepsilon^{2}=z(1-z)Q^{2}+m_{f}^{2}$,
$m_{f}$ is the quark mass and $z$ is the
fraction of photon's light-cone
momentum $q_{-}$ carried by one of the quarks of the pair
($0 <z<1$).
The flavour and $Q^{2}$ dependence of structure functions
is concentrated in
wave functions (\ref{eq:6},\ref{eq:7}), whereas the
dipole cross section $\sigma(x,r)$ is universal for all
flavours.
We emphasize that the factorization of the integrands in
Eq.~(\ref{eq:5}) follows from an {\sl exact}
diagonalization of the diffraction scattering matrix in the
$(\vec{r},z)$-representation [5,6]. Furthermore, the dipole-cross
section representation (\ref{eq:5}) and wave functions
(\ref{eq:6},\ref{eq:7}) are valid in the BFKL regime,
i.e., also beyond the Leading-Log$Q^{2}$ approximation
(LLQA). The $x$ (energy) dependence of the
dipole cross section $\sigma(x,r)$ comes from the higher
$q\bar{q}g_{1}...g_{n}$ Fock states of the photon, i.e., from
the QCD evolution effects [6] described by the
generalized BFKL equation. The transverse and
longitudinal structure functions are given by the familar
equation
$
F_{T,L}(x,Q^{2})=(Q^{2}/4\pi \alpha_{em})
\sigma_{T,L}(x,Q^{2}).
$
We advocate using $F_{T}(x,Q^{2})=2xF_{1}(x,Q^{2})$, because it
has simpler interpretation than $F_{2}=F_{T}+F_{L}$, which mixes
interactions of the transverse and longitudinal photons.

The ratio $\sigma(x,r)/r^{2}$ is a smooth function of $r$, and
it is convenient to use the representation
\beq
F_{T}(x,Q^{2}) = {1\over \pi^{3}} \int {dr^{2}\over r^{2}}
{\sigma(x,r)\over r^{2}}
\sum e_{f}^{2}\Phi_{T}^{(f\bar{f})}(Q^{2},r^{2})\, ,
\label{eq:8}
\endeq
\beq
F_{L}(x,Q^{2}) = {1\over \pi^{3}} \int {dr^{2}\over r^{2}}
{\sigma(x,r)\over r^{2}}
\sum e_{f}^{2}W_{L}^{(f\bar{f})}(Q^{2},r^{2})\, ,
\label{eq:9}
\endeq
\beq
{\partial F_{T}(x,Q^{2})\over \partial \log Q^{2}} =
{1\over \pi^{3}} \int {dr^{2}\over r^{2}}
{\sigma(x,r)\over r^{2}}
\sum e_{f}^{2}W_{T}^{(f\bar{f})}(Q^{2},r^{2})\, ,
\label{eq:10}
\endeq
where the weight functions $\Phi_{T}^{(f\bar{f})}$ and
$W_{T,L}^{(f\bar{f})}$ are defined by
\beq
\Phi_{T}^{(f\bar{f})}(Q^{2},r^{2}) =
(\pi^{2}/4\alpha_{em})\int_{0}^{1} dz \, Q^{2}r^{4}
|\Psi_{T}^{(f\bar{f})}(z,r)|^{2}\, ,
\label{eq:11}
\endeq
\beq
W_{L}^{(f\bar{f})}(Q^{2},r^{2}) =
(\pi^{2}/4\alpha_{em})\int_{0}^{1} dz \, Q^{2}r^{4}
|\Psi_{L}^{(f\bar{f})}(z,r)|^{2}\,
\label{eq:12}
\endeq
\beq
W_{T}^{(f\bar{f})}(Q^{2},r^{2}) =
{\partial \Phi_{T}^{(f\bar{f})}(Q^{2},r^{2})\over
\partial \log Q^{2}}\,.
\label{eq:13}
\endeq
These weight functions ar shown in Figs.~1,2. Following the
above outlined strategy, we wish
identify the observable $f(x)$ which is dominated by the
contribution from $r=r_{\Delta}$, so that its $x$-dependence can
be used to determine the pomeron's intercept $\Delta_{\Pom}=
-\log f(x)/d\log x$.
Fig.~1 shows that the transverse structure function
$F_{T}(x,Q^{2})$ receives contributions from
$
{B_{T}/ Q^{2}} \lsim r^{2} \lsim {1/ m_{f}^{2}}\,,
$
does not zoom at $r\sim r_{\Delta}$, and as such it is not suited
for determination of $\Delta_{\Pom}$.

Let us discuss the salient features of
$\Phi_{T}^{(f\bar{f})}(Q^{2},r^{2})$ in more detail.
At large $Q^{2}$ it developes a plateau of unit height.
The width of the plateau rises
$\propto \log Q^{2}$, and the emergence of this plateau signals
the onset of the familiar
LLQA. At $r^{2} \gg B_{T}/Q^{2}$, the weight function
$\Phi_{T}^{(f\bar{f})}(Q^{2},r^{2})$ is a scaling function of
$Q^{2}$ and, because $\sigma(x,r)$ does not
depend on $Q^{2}$, the $Q^{2}$-dependence of
$F_{T}(x,Q^{2})$ entirely comes
from the expansion of the plateau of
$\Phi_{T}^{(f\bar{f})}(Q^{2},r^{2})$ towards small $r$
with increasing $Q^{2}$.

Notice a very slow onset of the scaling regime
for the charmed quark contribution. The corresponding
$\Phi_{T}^{(f\bar{f})}(Q^{2},r^{2})$ starts developing a plateau
only at very large $Q^{2}$. Fig.~1 strongly suggests that
the charmed quarks can be treated as massles partons, the
charm can be regarded an active flavour and the LLQA becomes
accurate for the charmed quarks, only at
$Q^{2} \gsim 100$GeV$^{2}\sim 40 m_{c}^{2}$.
At very small sizes, $Q^{2}r^{2}\ll B_{T}$ and $r^{2}m_{c}^{2}\ll 1$,
the weight functions for the charmed and light quarks converge
to each other already at moderate $Q^{2}$.
For the light $u,d$ quarks, the curves shown are for
$m_{u,d}=0.15$GeV. In this case the onset of the plateau, and
of the LLQA thereof, requires $Q^{2}\gsim (2-3)$GeV$^{2}$.
(With the more conservative $m_{u,d}=0.3$GeV the plateau only starts
developing, and the LLQA only will be accurate, at
$Q^{2}\gsim 10$GeV$^{2}$). In Fig.~1 we also show
$$
\Phi_{T}(Q^{2},r^{2})= {9\over 11}
\sum_{u,d,s,c,b} e_{f}^{2} \Phi_{T}^{(f\bar{f})}(Q^{2},r^{2})\, ,
$$
which would have been equal to unity if all 5 flavours were active.
The large mass of the charmed quark has a very profound effect,
and it is evidently premature to speak of the
4 active flavours unless $Q^{2} \gsim (100-200)$GeV$^{2}$.

Evidently, $W_{T}^{(f\bar{f})}(Q^{2},r^{2}) =
\partial \Phi_{T}^{(f\bar{f})}(Q^{2},r^{2})/\partial \log Q^{2}$
will be a sharply peaked function of $r^{2}$, which only is
nonvanishing at $r^{2} \sim 1/Q^{2}$. In Fig.~2 we show
$W_{T,L}^{(f\bar{f})}(Q^{2},r^{2})$ as a function of the natural
variable
$\log(Q^{2}r^{2})$. At large $Q^{2}$ and $Q^{2}r^{2}\sim 1$,
$W_{T}^{(f\bar{f})}(Q^{2},r^{2})$ becomes a scaling function
of $Q^{2}$, in agreement with the conventional expectation that
in DIS the only relevant scale is $1/\sqrt{Q^{2}}$. At the
asymptotically large $Q^{2}$, the charmed and light quarks have
identical $W_{T}(Q^{2},r^{2})$. The $Q^{2}$, $r^{2}$ and the
flavour dependence of $W_{L}(Q^{2},r^{2})$ resembles
that of $W_{T}(Q^{2},r^{2})$, apart from the strikingly slower
onset of the asymptotic scaling form of
$W_{L}(Q^{2},r^{2})$ and, consequently, of the LLQA for the
longitudinal structure function.

The weight functions $W_{T,L}(Q^{2},r^{2})$
have their center of gravity at $Q^{2}r^{2}=B_{T,L}$,
where $B_{T}\approx 2.1,\,2.3,\, 2.6,\, 2.8$
and $B_{L}\approx 8.3,\, 10.2,\, 10.6,\,12$ at
$Q^{2}= 0.75,\, 4.5,\, 30$ and $480$GeV$^{2}$, respectively.
Therefore, at large $Q^{2}$ (summation only over active
flavours is understood)
\beq
F_{L}(x,Q^{2}) = {1\over \pi^{3}}
\sum e_{f}^{2}
\left.{\sigma(x,r)\over r^{2}}\right|_{r^{2}=B_{L}/Q^{2}}\,\, ,
\label{eq:15}
\endeq
\beq
{\partial F_{T}(x,Q^{2})\over \partial \log Q^{2}} =
{1\over \pi^{3}}
\sum e_{f}^{2}
\left.{\sigma(x,r)\over r^{2}}\right|_{r^{2}=B_{T}/Q^{2}}\,\, ,
\label{eq:16}
\endeq
and the $x$ dependence of $F_{L}(x,Q^{2})$ and
$\partial F_{T}(x,Q^{2})/ \partial \log Q^{2}$ at values of
$Q^{2}$ given by equations (\ref{eq:3}) and (\ref{eq:4}),
respectively, follows a precocious asymptotic behaviour of
$\sigma(x,r_{\Delta})$ and
measures the pomeron's intercept.

Fig.~1 shows that, by the numerical coincidence $r_{\Delta}\sim
1/m_{c}$,
at $Q^{2} \lsim 10$GeV$^{2}$ the charm
contribution to $F_{T}(x,Q^{2})$ is also dominated by the
contribution from $r\sim r_{\Delta}$. Therefore, as we
suggested in [2], the measurement of the $x$-dependence
of the charm structure function $F_{T}^{(c\bar{c})}(x,Q^{2})$
also allows determination of $\Delta_{\Pom}$. The
availability of the above three methods of determination of
the pomeron's intercept is very important as allows the
consistency checks.

Two important points, common to all the above methods of
determination of $\Delta_{\Pom}$, must be emphasized.
Firstly, they are quite insensitive to
the belated onset of LLQA and to the precise number of
active flavours (ss a matter of fact, the excitation of
charm at $Q^{2}\lsim 10$GeV$^{2}$
corresponds to a deeply sub-LLQA regime). The suggested
technique only depends on
the position  of the peak of $W_{T,L}(Q^{2},r^{2})$, the
existence of this peak does not require the applicability
of LLQA. Secondly, it would have been somewhat more
accurate to substitute $\sigma(x,r)$ in the integrand of
(\ref{eq:5}) for $\sigma({x\over z},r)$. However, as
far as we zoom at $r\approx r_{\Delta}$, at which
$\sigma(x,r_{\Delta}) \propto \left({1\over x}\right)
^{\Delta_{\Pom}}$, this substitution does not affect the
$x$-dependence in Eqs.~(\ref{eq:15},\ref{eq:16}) and the
proposed determination
of the pomeron's intercept $\Delta_{\Pom}$. With the advent
of the high precision data, the above approach can readily
be extended  to include the effects of the $z$-dependence of
the effective energy in the dipole cross section $\sigma(x,r)$,
and also the effects of the $z$-$r$ correlations in the
corresponding wave functions.

The comment on uncertainties and challenges is in order.
Much depends on the specific value of $r_{\Delta}$.
Our analysis [2,3] has shown that $r_{\Delta}
\sim {1\over 2}R_{c}$, where $R_{c}$ is the correlation
radius for the perturbative gluons. This correlation
radius was a subject of active studies in the lattice QCD,
which suggest $R_{c}\approx 0.3$f (for the review
and references see [8]).
The analysis [2,4,6] also strongly suggests that the existence
of the magic dipole size $r_{\Delta}$ at which the dipole cross
section exhibits a precocious asymptotic behaviour,
is quite a generic property of the BFKL
equation. However, besides the gluon correlation radius
$R_{c}$, there may be other nonperturbative parameters which
can affect both the pomeron's intercept and the magic size
$r_{\Delta}$. Here we we only wish to notice that,
if $R_{c}\sim 0.3$f as the lattice QCD studies
suggest, then $r_{\Delta} \sim {1\over 2}R_{c}$
corresponds to distances at which
the QCD coupling is relatively small. Also,
the analysis [4] suggests that at
so small a value of $r_{\Delta}$ the
nonperturbative component of the dipole cross section can be
neglected.
For this reasons, one can anticipate
weak dependence of $r_{\Delta}$ on the nonperturbative effects
at $r\gsim 1$f. More studies on the nonperturbative effects
in the BFKL eqation are evidently needed.

There are two byproducts of the above analysis, both of
potential significance for determination [9,10] of the gluon
structure function $G(x,Q^{2})$ from $F_{L}(x,Q^{2})$ and
$\partial F_{T}(x,Q^{2})/\partial \log Q^{2}$:
a very slow onset of the scaling regime
for the charmed quark contribution to structure functions
and the subtantial difference between
$B_{T}$ and $B_{L}$.  The former is important for the
number of active favours, the latter suggests that
$F_{L}(x,Q^{2})$ and
$\partial F_{T}(x,Q^{2})/\partial \log Q^{2}$ will give
$G(x,q^{2})$ at different values of $q^{2}$,
which differ by a large factor $B_{L}/B_{T}$, and which are
both different from $Q^{2}$. The anomalously large
value of $B_{L}$ suggests a very slow onset of the short-distance
dominance, and the potentially large corrections to LLQA, for
the longitudinal structure function. The very slow onset of
LLQA for charmed quarks affects a number of active flavours and
is still another potential source  of large
corrections to the much discussed LLQA relations [9,10] between
$G(x,Q^{2})$ and $F_{L}(x,Q^{2}), \,
\partial F_{T}(x,Q^{2})/\partial \log Q^{2}$. The
detailed discussion of these corrections to LLQA will be
presented elsewhere [11].\medskip\\
\centerline{\bf \large Conclusions:}

The purpose of this study has been an analysis of determination
of the pomeron's intercept $\Delta_{\Pom}$
from measurements of the longitudinal
$F_{L}(x,Q^{2})$, and the slope $\partial F_{T}(x,Q^{2})
/\partial \log Q^{2}$ of the transverse, structure functions
at HERA. Our principal finding is that $\Delta_{\Pom}$
can readily be determined at HERA concentrating on the special
range of $Q^{2}$ given by Eqs.~(\ref{eq:3},\ref{eq:4}).
The $x$-dependence of the
excitation of charm at $Q^{2}\lsim 10$GeV$^{2}$ provides the
third, independent, determination of $\Delta_{\Pom}$. The
important virtue of proposed methods is their
model-independence: they require neither a validity of
the Leading-Log$Q^{2}$ approximation, nor a knowledge of the
number of active flavours. They require, though, a knowledge
of the magic dipole size $r_{\Delta}$, which we related to
the nonperturbative correlation radius of the perturbative
gluons, known from the lattice QCD studies. Still, in
parallel with the anticipated experimental determinations of
$\Delta_{\Pom}$, more work
on the noneprturbative effects in the BFKL equation is needed.
\medskip\\
\centerline{\bf \large Acknowledgements}
One ot the authors (BGZ) would like to thank F.Close, J.Forshaw
and R.Roberts for useful discussions on the longitudinal
structure function and gluon distributions, J.Speth for
the hospitality extended at Inst.f.Kernphysik, KFA J\"ulich,
where this work started and S.Fantoni for hospitality at ILAS,
Trieste.
\pagebreak

\pagebreak
{\bf \Large Figure captions}
\begin{itemize}
\item[Fig.1]
 - The weight function $\Phi_{T}(Q^{2},r^{2}$: (a) for the light
flavours $u$ and $d$, (b) for the charmed quark, (c) for the
$b$-quark, (d) the global weight function for 5 flavours
$(u,d,s,c,b)$. The dashed, dotted, solid, lohg-dashed and
dot-dashed curves are for $Q^{2}=0.75,\, 4.5$,\, 30,\, 240,\,
$2000$GeV$^{2}$, respectively.

\item[Fig.2]
 - The weight functions $W_{T,L}(Q^{2},r^{2})$: (a) the light
flavours $(u,d)$, (b) the charmed quark, (c) the global weight
function for 5 flavours $(u,d,s,s,b)$. The dashed, dotted,
dot-dashed and solid curves are for $Q^{2}=0.75,\,4.5,\,30,
\,240$GeV$^{2}$, respectively.

\end{itemize}
\end{document}